\documentclass{PoS}
\newcommand{\be}{\begin{equation}}
	\newcommand{\ee}{\end{equation}}
\newcommand{\bea}{\begin{eqnarray}}
	\newcommand{\eea}{\end{eqnarray}}
\newcommand{\ba}{\begin{array}}
	\newcommand{\ea}{\end{array}}

\usepackage{graphicx}
\usepackage{epstopdf}
\usepackage{amsmath}
\usepackage{slashed}

\title{Pion wavefunction with dynamical spin effects}

\ShortTitle{Pion wavefunction with dynamical spin effects}

\author{\speaker{Mohammad Ahmady}%
        \\
        Department of Physics, Mount Allison University, Sackville, New Brunswick, Canada E4L 1E6\\
        E-mail: \email{mahmady@mta.ca}}

\author{Farrukh Chishtie
	\\ 
	Department of Physical Sciences,
	Theoretical Research Institute, Pakistan Academy of Sciences (TRIPAS), Islamabad 44000, Pakistan\\ 
	E-mail: \email{fchishti@uwo.ca}}

\author{Ruben Sandapen 
	\\ 
	Department of Physics, Acadia University,
	Wolfville, Nova-Scotia, Canada, B4P 2R6 and
	Department of Physics, Mount Allison University,
	Sackville, New Brunswick, Canada, E4L 1E6\\ 
	E-mail: \email{ruben.sandapen@acadiau.ca}} 

\abstract{We report that the inclusion of dynamical spin effects in the pion holographic light-front wavefunction leads to a remarkable improvement in describing pion observables (pion mean charge radius, decay constant, spacelike electromagnetic form factor) without the need to invoke higher Fock state contributions and/or a special AdS/QCD mass scale for the pion.}

\FullConference{XVII International Conference on Hadron Spectroscopy and Structure\\
		 25-29 September, 2017\\
		 University of Salamanca, Salamanca, Spain}

\begin{document}

\section{Introduction}
 The hadronic light-front wavefunctions (LFWFs) are obtained by solving the LF Heisenberg equation for QCD: \cite{Brodsky:2014yha}
\begin{equation}
	H_{\text{QCD}}^{\text{LF}} |\Psi(P)\rangle = M^2 |\Psi(P) \rangle\;,
	\label{LFQCD} 
\end{equation} 
where $H_{\text{QCD}}^{\text{LF}}=P^+P^- -P_{\perp}^2$ and $M$ is the hadron mass. At equal light-front time $(x^+=0)$ and in the light-front gauge $A^+=0$, the hadron state $|\Psi(P)\rangle$ is expanded in terms of the Fock states:  
\begin{equation}
	|\Psi(P^+, \mathbf{P_{\perp}}, S_z) \rangle = \sum_{n,h_i} \int [\mathrm{d} x_i]  [\mathrm{d}^2 \mathbf{k}_{\perp i}] \frac{1}{\sqrt{x_i}}\Psi_{n}(x_i,\mathbf{k}_{\perp i},h_i) |n: x_i P^+, x_i \mathbf{P_{\perp}} + \mathbf{k}_{\perp i}, h_i \rangle\;,
	\label{Fock-expansion}	
\end{equation} 
where $\Psi_{n}(x_i, \mathbf{k}_{\perp i},h_i)$ is the LFWF of the  Fock state with $n$ constituents and 
\begin{equation}
	[\mathrm{d} x_i] \equiv \prod_{i}^{n} \mathrm{d} x_i \delta (1-\sum_{j=1}^{n} x_j) \hspace{1cm} [\mathrm{d}^2 \mathbf{k}_{i}]\equiv \prod_{i=1}^{n} \frac{\mathrm{d}^2 \mathbf{k}_i}{2(2\pi)^3} 16 \pi^3 \delta^2(\sum_{j=1}^n \mathbf{k}_i) \;,
	\label{Int-measures}	 
\end{equation} 
are the integration measures.  $x_i=k_i^+/P^+$ are the momenta fractions, $\mathbf{k}_{\perp i}$ represent the transverse momenta, and $h_i$ are the helicities of the constituents. To solve Eq. \eqref{LFQCD} the semiclassical approximation in which quark masses and quantum loops are neglected is assumed and therefore the LFWFs depend on the invariant mass $\mathcal{M}^2=(\sum_{i}^n k_i)^2$ of the constituents rather than on their individual momenta $k_i$. For mesons, the invariant mass of the $q\bar{q}$ pair in the valence ($n=2$) Fock state is $\mathcal{M}_{q\bar{q}}^2=k_{\perp}^2/x(1-x)$ which is the Fourier conjugate to the impact variable $\zeta^2=x(1-x) b^2$ where $b$ is the transverse separation of the quark and antiquark \cite{Brodsky:2014yha}. Doing so, the valence meson LFWF can then be factorized as: 
\begin{equation}
	\Psi(\zeta, x, \phi)= e^{iL\phi} \mathcal{X}(x) \frac{\phi (\zeta)}{\sqrt{2 \pi \zeta}} \;.
	\label{mesonwf}
\end{equation}
This formulation of the LFWF assumes that the helicity dependence decouples from the dynamics.  As a result, one can show that the transverse component of Eq. \eqref{mesonwf}, i.e. $\phi (\zeta)$, is the solution of a $1$-dimensional Schr\"odinger-like wave equation, namely: 
\begin{equation}
	\left(-\frac{d^2}{d\zeta^2} - \frac{1-4L^2}{4\zeta^2} + U(\zeta) \right) \phi(\zeta) = M^2 \phi(\zeta) \;.
	\label{hSE}
\end{equation} 
The confinement potential $U(\zeta)$, which is yet to be determined, contains the interaction terms and the effects of higher Fock states on the valence state.  Indeed, Brodsky and de T\'eramond \cite{deTeramond:2008ht,Brodsky:2006uqa,deTeramond:2005su,Brodsky:2007hb} found that Eq. \eqref{hSE} maps onto the wave equation for the propagation of spin-$J$ string modes in the higher dimensional anti-de Sitter space, $\text{AdS}_5$.  In this correspondence, $\zeta$ is identified with $z_5$, the fifth dimension of AdS space and the light-front orbital angular momentum $L^2$ is mapped onto $(mR)^2-(2-J)^2$  where $R$ and $m$ are the AdS radius and mass respectively. Hence, the Eq. \eqref{hSE} is referred to as the holographic LF Schr\"odinger equation. In this AdS/QCD duality, the confining potential is related to the dilaton field $\varphi(z_5)$, which breaks the conformal symmetry in AdS, as follows:
\begin{equation}
	U(z_5, J)= \frac{1}{2} \varphi^{\prime\prime}(z_5) + \frac{1}{4} \varphi^{\prime}(z_5)^2 + \left(\frac{2J-3}{4 z_5} \right)\varphi^{\prime} (z_5) \;.
\end{equation}
Assuming quadratic dilaton profile, i.e.  $\varphi(z_5)=\kappa^2 z_5^2$, leads to a light-front harmonic oscillator potential in physical spacetime:
\begin{equation}
	U(\zeta,J)= \kappa^4 \zeta^2 + \kappa^2 (J-1) \;.
	\label{harmonic-LF}
\end{equation}
keeping in mind that $z_5$ maps onto the LF impact variable $\zeta$. Indeed, the quadratic form of the AdS/QCD potential is unique as shown in Ref \cite{Brodsky:2013ar}.

Inserting the confining potential given in \eqref{harmonic-LF}, one can obtain the eigenvalues,
\begin{equation}
	M^2= 4\kappa^2 \left(n+L +\frac{S}{2}\right)\;,
	\label{mass-Regge}
\end{equation}
and their corresponding eigenfunctions
\begin{equation}
	\phi_{nL}(\zeta)= \kappa^{1+L} \sqrt{\frac{2 n !}{(n+L)!}} \zeta^{1/2+L} \exp{\left(\frac{-\kappa^2 \zeta^2}{2}\right)} L_n^L(x^2 \zeta^2) \;,
	\label{phi-zeta}
\end{equation}
of the holographic Sch\" odinger equation \eqref{hSE}.  We note that the lowest meson state, i.e. $n=L=S=0$, which is identified with pions, is massless as is expected and meson masses lie on linear Regge trajectories ($M^2\sim L$).  The longitudinal mode $\mathcal{X}(x)$ of the meson LF wavefunction \eqref{hwf} is obtained by matching the expressions for the pion EM or gravitational form factor in physical spacetime and in AdS space. One obtains $\mathcal{X}(x)=\sqrt{x(1-x)}$ \cite{deTeramond:2008ht,Brodsky:2008pf} through this matching technique. Therefore,  in the massless quark limit, meson holographic LFWFs can be written as:
\begin{equation}
	\Psi_{nL}(\zeta, x, \phi)= e^{iL\phi} \sqrt{x(1-x)} (2\pi)^{-1/2}\kappa^{1+L} \sqrt{\frac{2 n !}{(n+L)!}} \zeta^{L} \exp{\left(\frac{-\kappa^2 \zeta^2}{2}\right)} L_n^L(x^2 \zeta^2)\;.
\end{equation}
To make connection with experiment, it is necessary to restore both the quark mass and helicity dependence of the holographic LFWF. Indeed, non-zero quark masses is usually implemented following the prescription of Brodsky and de T\'eramond \cite{Brodsky:2008pg}. For the ground state pion, this leads to 
\be  
\Psi^{\pi} (x,\zeta^2) = \mathcal{N} \sqrt{x (1-x)}  \exp{ \left[ -{ \kappa^2 \zeta^2  \over 2} \right] }
\exp{ \left[ -{m_f^2 \over 2 \kappa^2 x(1-x) } \right]}\;.
\label{hwf}
\ee
The normalization constant $\mathcal{N}$ is fixed by requiring that
\begin{equation}
	\int \mathrm{d}^2 \mathbf{b} \mathrm{d} x |\Psi^{\pi}(x,\zeta^2)|^2 = P_{q\bar{q}} 
	\label{norm}
\end{equation}
where $P_{q\bar{q}}$ is the probability that the meson consists of the leading quark-antiquark Fock state.  

Note that Eq. \eqref{mass-Regge} tells us that the AdS/QCD scale $\kappa$ can be chosen to fit the experimentally measured Regge slopes. Ref. \cite{Brodsky:2014yha} reports $\kappa=590$ MeV for pseudoscalar mesons and $\kappa=540$ MeV for vector mesons.  A recent fit to the  Regge slopes of mesons and baryons, treated as conformal superpartners, yields  $\kappa=523$ MeV \cite{Brodsky:2016rvj}. On the other hand, the AdS/QCD scale $\kappa$ can be connected to the scheme-dependent pQCD renormalization scale $\Lambda_{\text{QCD}}$ by matching the running strong coupling in the non-perturbative (described by light-front holography) and the perturbative regimes \cite{Brodsky:2010ur}. With $\kappa=523$ MeV and the $\beta$-function of the QCD running coupling at $5$-loops,  Brodsky, Deur and de T\'eramond recently predicted the QCD renormalization scale, $\Lambda^{{\overline{MS}}}_{\text{QCD}}$, in excellent agreement with the world average value \cite{Deur:2016opc}. Furthermore, light-front holographic wavefunctions have also been used to predict diffractive vector meson production\cite{Forshaw:2012im,Ahmady:2016ujw}. A fit to the HERA data on diffractive $\rho$ electroproduction, with $m_{u/d}=140$ MeV, gives $\kappa=560$ MeV\cite{Forshaw:2012im} and using $\kappa=550$ MeV (with $m_{u/d}[m_s]=46[140]$ MeV) leads to a good simultaneous description of the HERA data on diffractive $\rho$ and $\phi$ electroproduction \cite{Ahmady:2016ujw}. These findings hint towards the emergence of a  universal fundamental AdS/QCD scale $\kappa \sim 550$ MeV. In the most recent application of LFH to predict nucleon EM form factors \cite{Sufian:2016hwn}, it is pointed out that this universality holds up to $10\%$ accuracy. In this paper, we shall use the value of $\kappa=523$ MeV which fits the meson/baryon Regge slopes and accurately predicts $\Lambda^{{\overline{MS}}}_{\text{QCD}}$ \cite{Brodsky:2016rvj, Deur:2016opc}.

In earlier applications of LFH with massless quarks, much lower values of $\kappa$ were required to fit the pion data: $\kappa=375$ MeV in Ref. \cite{Brodsky:2007hb} in order to fit the pion EM form factor data and $\kappa=432$ MeV (with $P_{q\bar{q}}=0.5$) to fit the photon-to-pion transition form factor data simultaneously at large $Q^2$ and $Q^2=0$ (the latter is fixed by the $\pi^0 \to \gamma \gamma$ decay width) \cite{Brodsky:2011xx}.  Note that in Ref. \cite{Brodsky:2007hb}, the EM form factor is computed, both in the spacelike and timelike regions, as a convolution of normalizable hadronic modes with a non-normalizable EM current which propagates in the modified infrared region of AdS space and generates the non-perturbative pole structure of the EM form factor in the timelike region. Alternatively, the spacelike EM form factor can be computed using the Drell-Yan-West formula \cite{Drell:1969km,West:1970av} in physical spacetime with the holographic pion LFWF. The latter approach is taken in Refs. \cite{Vega:2008te,Swarnkar:2015osa,Vega:2009zb,Branz:2010ub}. In  Ref. \cite{Vega:2009zb}, a higher value of $\kappa=787$ MeV is used with $m_{u/d}=330$ MeV and the authors predict $P_{q\bar{q}}=0.279$, implying an important contribution of higher Fock states in the pion. In Ref. \cite{Branz:2010ub}, a universal AdS/QCD scale $\kappa=550$ MeV is used for all mesons, together with a constituent quark mass $m_{u/d}=420$ MeV, but $P_{q\bar{q}}=0.6$ is fixed for the pion only: for the kaon, $P_{q\bar{q}}=0.8$ and for all other mesons, $P_{q\bar{q}}=1$. More recently, in Ref. \cite{Swarnkar:2015osa}, with $m_{u/d}=330$ MeV, the authors use a universal $\kappa=550$ MeV for all mesons but fix the wavefunction normalization for the pion so as to fit the decay constant. Consequently, this implies that $P_{q\bar{q}}=0.61$ only for the pion.

All these previous studies seem to indicate that a special treatment is required at least for the pion either by using a distinct AdS/QCD scale $\kappa$ or/and relaxing the normalization condition on the holographic wavefunction, i.e. invoking higher Fock states contributions. This may well be reasonable since the pion is indeed unnaturally light and does not lie on a Regge trajectory, as pointed out in Ref. \cite{Vega:2008te}. However, we note that in the previous studies \cite{Swarnkar:2015osa,Vega:2009zb,Branz:2010ub} where the pion observables are predicted using the holographic wavefunction, given by Eq. \eqref{hwf}, the helicity dependence of the latter is always assumed to decouple from the dynamics, i.e. the helicity wavefunction is taken to be momentum-independent. This is actually consistent with the semi-classical approximation within which the AdS/QCD correspondence is exact. Consequently, Ref. \cite{Branz:2010ub} derives a single formula to predict simultaneously the vector and pseudoscalar meson decay constants, so that using a universal scale $\kappa$ and $P_{q\bar{q}}=1$ for all mesons inevitably leads to degenerate decay constants in conflict with experiment.

In this paper, we show that it is possible to achieve a better description of the pion observables by using a universal AdS/QCD scale $\kappa$ and without the need to invoke higher Fock state contributions. We do so by taking into account dynamical spin effects in the holographic pion wavefunction, i.e. we use a momentum-dependent helicity wavefunction.   This approach goes beyond the semiclassical approximation, just like the inclusion of light quark masses in the holographic wavefunction. However, it does support the idea of the emergence of a universal, fundamental AdS/QCD scale. A similar approach was taken previously for the $\rho$ meson, leading to impressive agreement to the HERA data on diffractive $\rho$ electroproduction \cite{Forshaw:2012im}.


\bibliographystyle{JHEP}
\bibliography{Hadron2017}

\end{document}